\documentclass[aps,twocolumn,pra,superscriptaddress,amsmath,showpacs,tightenlines]{revtex4-1}
\usepackage{amssymb}
\usepackage{amsmath}
\usepackage{braket}
\usepackage{graphicx}
\usepackage{subfigure}
\usepackage{natbib}
\usepackage{epsfig}
\usepackage{amsfonts}
\usepackage{mathrsfs}
\usepackage{CJK}
\usepackage{xcolor}
\usepackage[toc,page,title,titletoc,header]{appendix}

\begin{document}
\title{Quantum state preparation and transfer based on the bound state in the doublon continuum}
\author{Xiaojun Zhang}
\affiliation{Center for Quantum Sciences and School of Physics, Northeast Normal University,
Changchun 130024, China}
\author{Xiang Guo}
\affiliation{Center for Quantum Sciences and School of Physics, Northeast Normal University,
Changchun 130024, China}
\author{Yan Zhang}
\affiliation{Center for Quantum Sciences and School of Physics, Northeast Normal University,
Changchun 130024, China}
\author{Xin Wang}
\affiliation{Institute of Theoretical Physics and School of Physics, Xi'an Jiaotong University,
Xi'an 710049, China}
\author{Haijun Xing}
\email{hjxing@nenu.edu.cn}
\affiliation{Center for Quantum Sciences and School of Physics, Northeast Normal University,
Changchun 130024, China}
\author{Zhihai Wang}
\email{wangzh761@nenu.edu.cn}
\affiliation{Center for Quantum Sciences and School of Physics, Northeast Normal University,
Changchun 130024, China}

\begin{abstract}
Bound states in the continuum (BICs) have attracted intense interest, yet their many-particle counterparts remain largely unexplored in waveguide quantum electrodynamics. We identify and characterize a bound state embedded in the doublon continuum (BIDC) that emerges when four atoms couple to a coupled-resonator waveguide with strong on-site interaction. Exploiting this interaction-enabled BIDC, we show that (i) a distant, four-atom entangled state can be prepared with high fidelity, and (ii) quantum entangled states can be coherently transferred between spatially separated nodes. Our results establish a scalable mechanism for multi-particle state generation and routing in waveguide platforms, opening a route to interaction-protected quantum communication with many-particle BICs.
\end{abstract}
\maketitle
\section{introduction}

The bound state in the continuum~(BIC) is a spatially localized state that resides within the radiation continuum in the frequency (energy) domain~\cite{bic1,bic2,bic3,bic4}. Owing to its broad utility-including low-threshold lasing~\cite{laser1,laser2,laser3}, enhanced nonlinear responses~\cite{nlr1,nlr2}, high-efficiency waveguiding~\cite{waveguiding1,waveguiding2,waveguiding3,waveguiding4}, and quantum sensing~\cite{sense1,sense2}, BIC physics has been extensively explored.

In the quantum information community, BIC-enabled quantum state preparation~(QSP) has recently been demonstrated~\cite{sp1,sp2}. Within waveguide QED, prior studies have focused mainly {on single-particle BICs~\cite{an16,kim23,lv23}} or on BICs that do not involve quantum emitters~\cite{bicli,bidc1,bidc2}. By contrast, multi-particle BICs arising from interference among emitters remain largely unexplored. It is therefore an open question whether such states exist and, if so, whether they can be harnessed to generate entanglement among larger ensembles of emitters.

On the other hand, quantum state transfer~(QST) between distant nodes plays a critical role in scalable quantum information processing~\cite{nature08,natp14,sp00,sp03,sp09,sp20,sp13,zhengpra,QIP1,QIP2,QIP3}. The earliest protocol coupled two quantum memories to the terminating ports of a waveguide~\cite{qst1997}. Building on this idea, a vast body of QST schemes has been developed with photons serving as the primary information carriers~\cite{qst1,qst2,qst3,qst4}. In strongly correlated systems, Ref.~\cite{wx} proposed using doublons as the mediating excitations for QST. Here, a doublon~\cite{wangprl,doublon1,doublon2} denotes a two-boson bound state localized on the same or neighboring lattice sites by strong attractive (or repulsive) interactions. The emitters can lead to bound states out of the doublon continuum~\cite{wxdoublon} in such strongly correlated system. These advances motivate us to ask whether a bound state embedded in the doublon continuum~(BIDC) can be applied to perform QST.

To address these two questions, we analyze a system of two atom pairs (four atoms) coupled to a coupled-resonator waveguide~(CRW) with strong on-site interactions, modeled within the Bose-Hubbard framework. The waveguide supports an interaction-induced doublon continuum that lies below the two-particle scattering band in the energy (frequency) domain. In the presence of the atoms, we further identify a BIDC in which the atoms are strongly entangled, enabling high-fidelity preparation of multi-atom entangled states. Moreover, by time-dependent engineering of the atom-waveguide coupling, we propose a QST protocol that goes beyond conventional adiabatic schemes: by allowing tunneling between the BIDC and the scattering continua, the required transfer time is reduced by about two orders of magnitude.

\section{model}
\begin{figure}
  \centering
  \includegraphics[width=7cm]{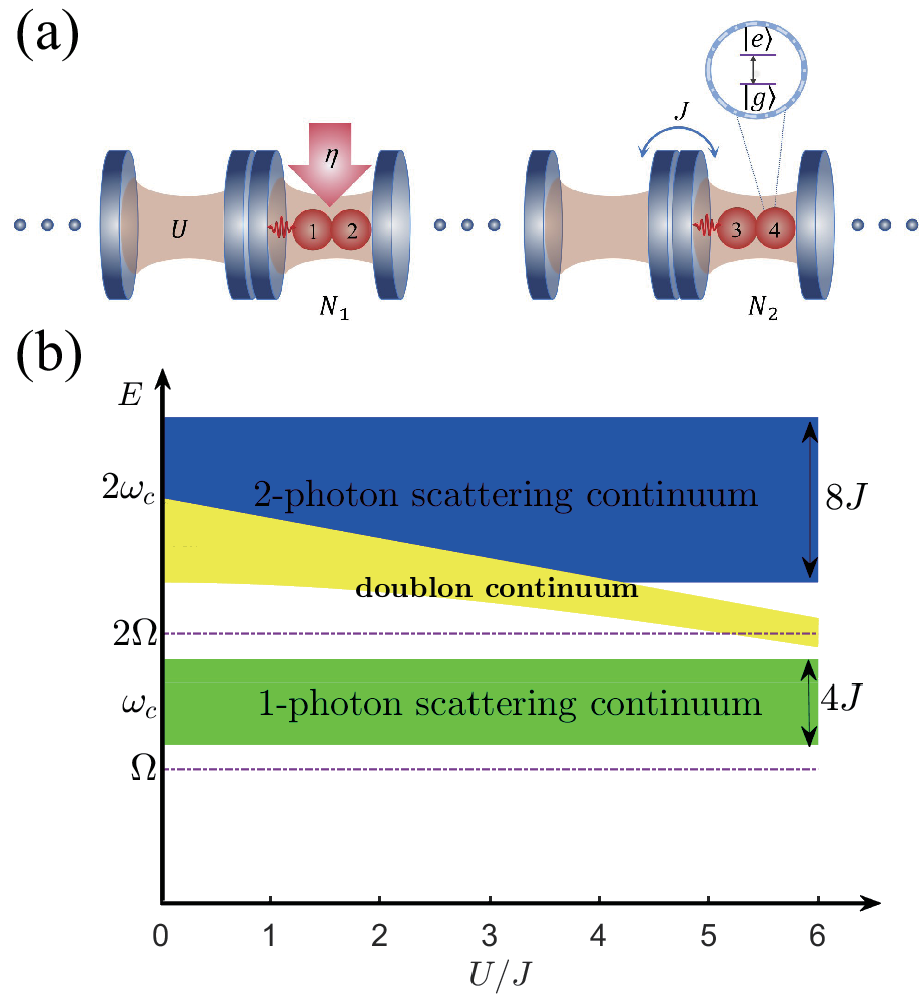}
  \caption{(a)~Sketch of the waveguide QED setup. (b)~The energy spectrum of the CRW. }\label{schem}
\end{figure}

We consider four two-level atoms coupled to the CRW, as shown in Fig.~\ref{schem}(a). The CRW Hamiltonian is
\begin{equation}
H_c=\sum_n \omega_ca_n^\dagger a_n-\frac{U}{2}\,a_n^\dagger a_n^\dagger a_n a_n
- J\left(a_n^\dagger a_{n+1}+\rm{H.c.}\right).
\end{equation}
where the characteristic frequencies of all resonators are the same $\omega_c$. \(U>0\) is the on-site interaction strength and \(J\) is the nearest-neighbor hopping amplitude. \(a_n\) (\(a_n^\dagger\)) annihilates (creates) a photon in the \(n\)th resonator.

The energy spectrum of the CRW is shown in Fig.~\ref{schem}(b). In the single-excitation subspace, the CRW hosts a single-photon scattering continuum of bandwidth \(4J\) with dispersion
\(\omega_k=\omega_c-2J\cos k\), where \(k\) is the photon wave vector. In the two-excitation subspace, the spectrum comprises both a two-photon scattering continuum (blue shaded region) and an interaction-induced doublon continuum (yellow shaded region). The doublon band has dispersion $\mathcal{E}_K=2\omega_c-\sqrt{U^2+16J^2\cos^2(K/2)}$~\cite{doublon1,doublon2},
where \(K\) is the center-of-mass (COM) wave vector of the biphoton. The corresponding eigenstates can be written as $|\psi_K^D\rangle=\sum_{m,n}{\rm exp}(iKx_c)\psi_{K}(r)a_m^\dagger a_n^\dagger|{\rm vac}\rangle/\sqrt{2N_c}$,
with \(x_c=(m+n)/2\) the COM coordinate, \(r=m-n\) the relative separation, \(N_c\) the number of resonators, and
$\psi_{K}(r)=\sqrt{\tanh(\lambda_K^{-1})}{\rm exp}({-|r|/\lambda_K})$
an exponentially localized wave function with $\lambda_K^{-1}={\rm{asinh}}(U_K)$ where $U_K=U/(4J\cos(K/2))$. In the doublon continuum, the two photons thus propagate as a correlated biphoton bound state within the waveguide. For simplicity, we shift the energy levels and set $\omega_c=0$ in the following.

The Hamiltonian of the whole structure is
\begin{align}
H=H_c+\sum_{i=1}^4\left[\Omega_i \sigma_i^+\sigma_i^-+g_i(\sigma_i^+a_{n_i}+\sigma_i^-a_{n_i}^\dagger)\right].\label{HA}
\end{align}
The second and third terms correspond to the free Hamiltonian of atoms and the interaction Hamiltonian between atoms and the CRW, respectively.
$\Omega_i$ denotes the $i$th atomic transition frequency between the ground state $|g\rangle$ and excited state $|e\rangle$, and $\sigma_i^+$~($\sigma_i^-$) is the raising~(lowering) operator. $g_i$ is the coupling strength between the $i$th atom and the CRW.

Such a model can be implemented in superconducting circuits. In particular, when the Josephson junction is replaced by a dc superconducting quantum interference device~(SQUID), the effective Josephson energy, and hence the transition frequency between the two qubit levels, can be tuned on demand by varying the magnetic flux threading the SQUID loop. Current experiments have demonstrated a tunability on the order of several hundred MHz~\cite{MD2017}. Therefore, it is feasible to engineer the transition frequencies of the four atoms as $\Omega_i=\Omega+\delta_i$, with $|\delta_i|\ll g_i$.

In this work, we take $g_1=g_2$ and $g_3=g_4$. Moreover, a photon-hopping strength of $J/(2\pi)=100$ MHz, which is also experimentally accessible and is adopted in our simulations below, can be achieved in superconducting-circuit platforms~\cite{J1}. Under this condition, the atomic transition frequency can be largely detuned from the single-photon scattering band, namely $E_{\mathrm{sl}}-\Omega\gg g_i$, where $E_{\mathrm{sl}}=\omega_c-2J$ denotes the lower band edge of the single-photon scattering continuum. In this parameter regime, the single-photon emission from an individually excited atom is effectively suppressed due to the large detuning. By contrast, two atoms can collectively emit a photonic doublon provided that $2\Omega\in\mathcal{E}_K$~\cite{wangprl,wx}, as illustrated in Fig.~\ref{schem}(b).

\begin{figure*}[t]
  \centering
  \includegraphics[width=15cm]{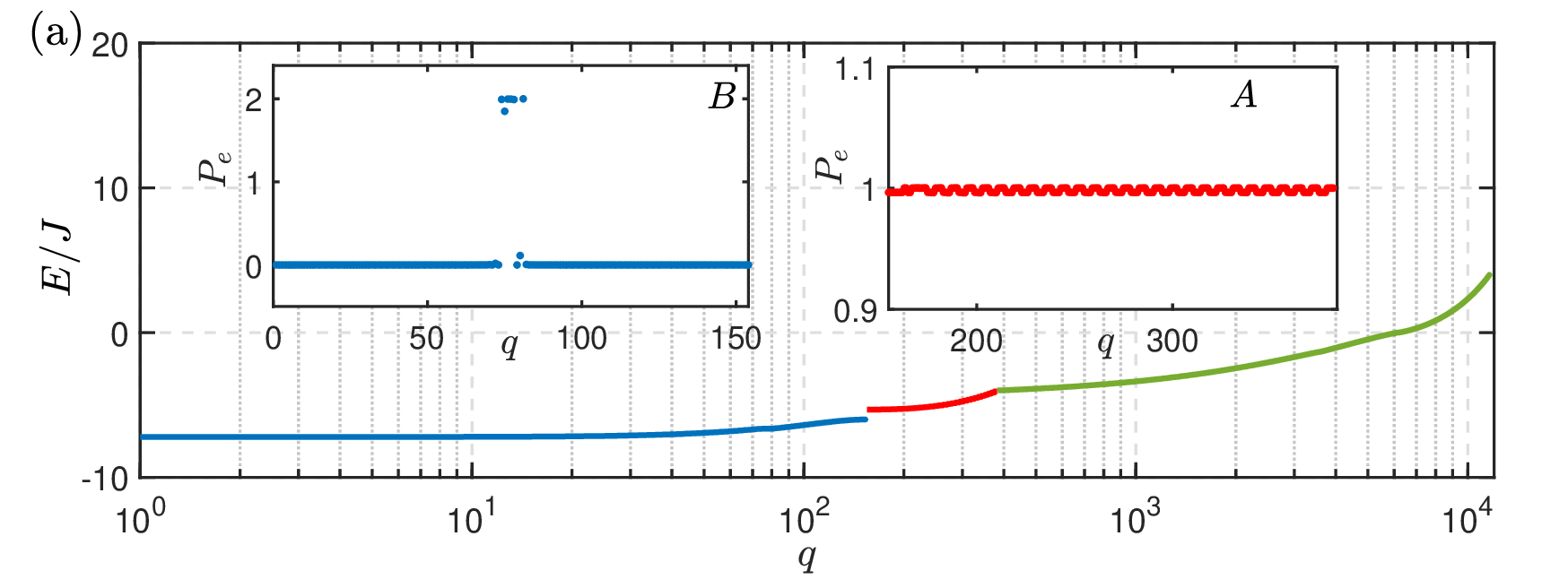}
  \includegraphics[width=15cm]{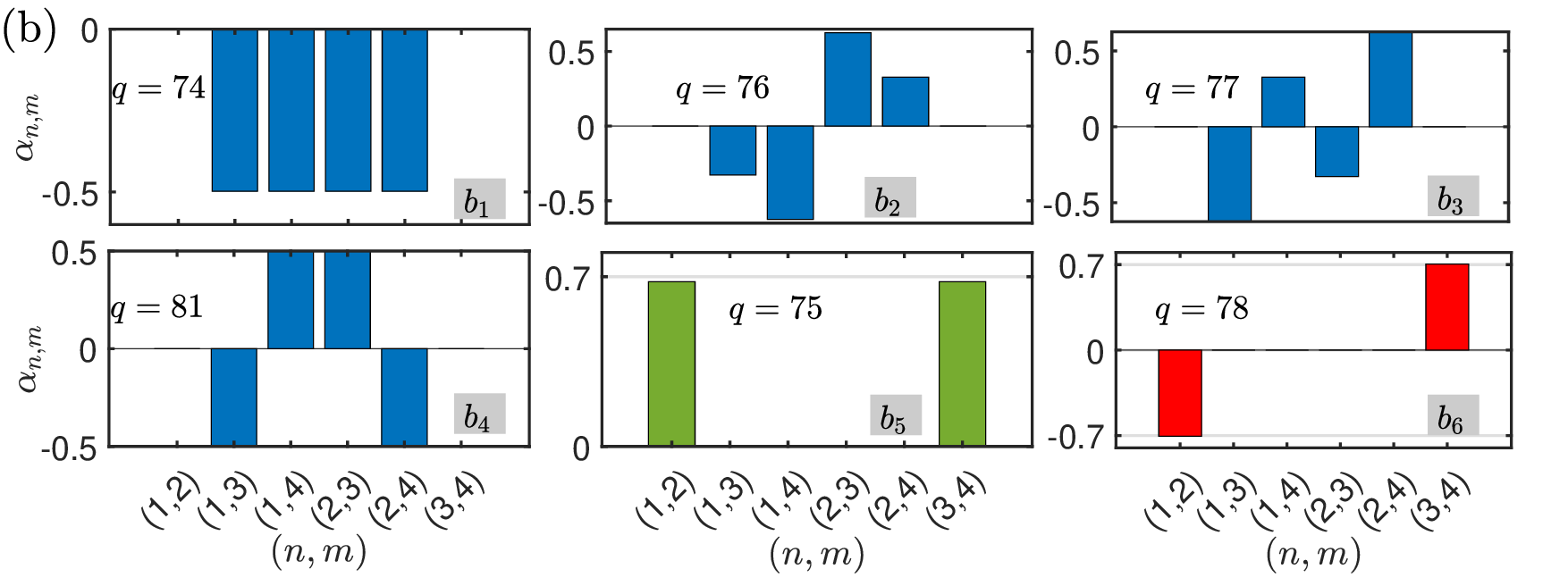}
  \includegraphics[width=16cm]{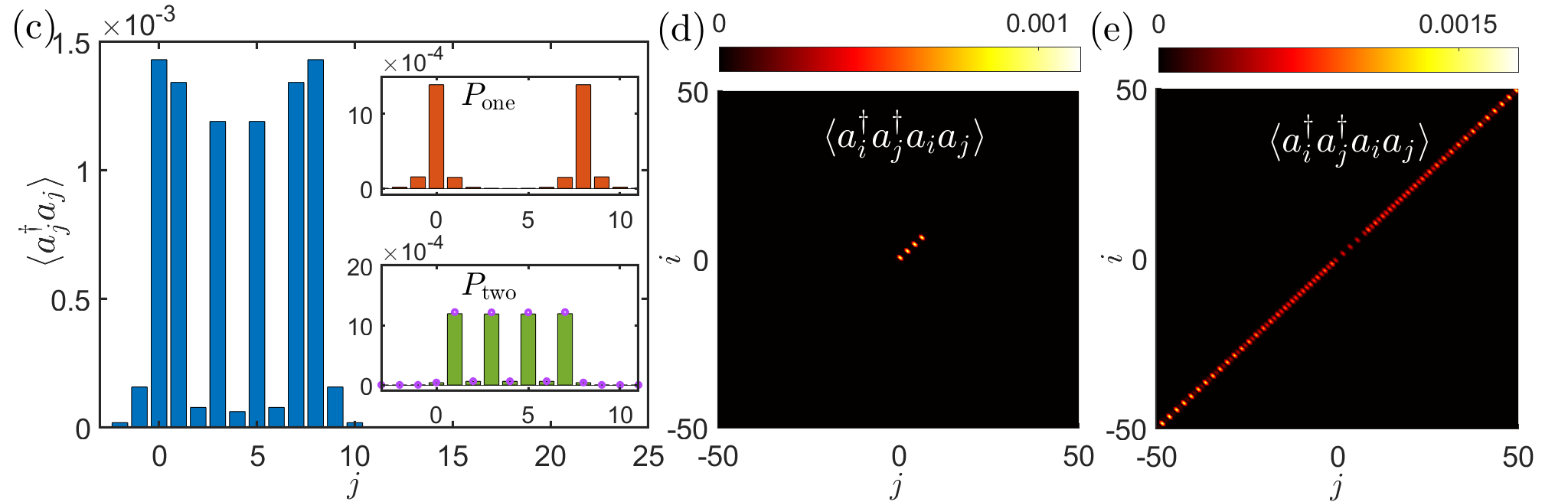}
  \caption{(a)~The eigenenergy of the whole structure. (b)~The atomic excitation probability amplitude. (c)~The photon distribution of the BIDC. The two panels correspond to the one-photon component $P_{\rm{one}}$ and the two-photon component $P_{\rm{two}}$, respectively. For $P_{\rm{two}}(j)$, the numerical results are represented by the green bars, while the results from Eq.~(\ref{cmn}) are shown by the purple circles. (d,e)~The two-photon probability distribution of the eigenstates $q=78$ and $q=75$.  (a-e)~The parameters $N_c=148$, $N_1=0,N_2=8$, $U=6J$, $\Omega_i=\mathcal{E}_{\pi/2}/2$ and $g_i=0.1J$~($i=1,2,3,4$). }\label{energy}
\end{figure*}

\section{bound state in the doublon continuum}
\subsection{Numerical results}
We first analyze the eigenenergies and eigenstates of the full system.
{Since} the total excitation number is conserved by Eq.~(\ref{HA}), the two-excitation sector is closed.
A general eigenstate \(\ket{\phi}\) in this sector can be expanded as
\begin{align}
|\phi\rangle=&\bigg[\sum_{i\leq j}c_{i,j}a_{i}^\dagger a_j^\dagger+\sum_{n< m}^{4}\alpha_{n,m} \sigma_n^+\sigma_m^+\nonumber\\
&+\sum_{n,i}d_{n,i}\sigma_n^+a_i^\dagger\bigg]|G,{\rm vac}\rangle.\label{eig}
\end{align}
where \(\ket{G,\mathrm{vac}}\) denotes all atoms in the ground state and the CRW in the vacuum state.
Here, \(c_{i,j}\), \(\alpha_{n,m}\), and \(d_{n,i}\) are expansion coefficients; \(i,j\) label resonator sites in the CRW, and \(n,m\in\{1,2,3,4\}\) label the atoms.

Under periodic boundary conditions, we numerically obtain the eigenenergies of the full system, as shown in Fig.~\ref{energy}(a), with the horizontal axis giving the eigenstate index \(q\).
The spectrum separates into three regions, highlighted by green, red, and blue curves.
The green branch corresponds to scattering states.
The red branch lies outside the two-photon scattering continuum and is therefore indicative of bound states.
To further characterize these states, we plot the total atomic excitation
\(
P_e=\sum_{i=1}^{4}\langle \sigma_i^+\sigma_i^-\rangle
\) in inset panel \(A\).
We find that approximately one excitation remains localized on the atoms, which identifies these eigenstates as single-photon bound states.

In this work we focus on the states indicated by the blue branch, where photons propagate as doublons.
Inset panel \(B\) shows the total atomic excitation \(P_e\) for these eigenstates;
for part of them \(P_e\simeq 2\), implying that the atomic sector carries nearly two excitations while only a small photonic component resides in the waveguide.
To gain further insight, Fig.~\ref{energy}(b) presents the atomic pair amplitudes \(\alpha_{n,m}\) for the state with $P_e\simeq 2$.
For clarity, we distinguish two classes of atomic pairs:
type-I atomic pair located in different resonators and
type-II atomic pair occupying the same resonator.

For the eigenstates in Figs.~\ref{energy}($b_1$-$b_4$), only type-I pairs are excited, that is, the amplitudes
\(\alpha_{1,3}\), \(\alpha_{1,4}\), \(\alpha_{2,3}\), and \(\alpha_{2,4}\) are nonzero.
Because the separation between the corresponding coupling sites is sufficiently large, the effective coupling between type-I pairs and the doublon continuum becomes negligibly small. Consequently, atomic decay is strongly suppressed (see Appendix~\ref{A} for details).
By contrast, for the eigenstates in Figs.~\ref{energy}($b_5$,$b_6$), only type-II pairs are populated, i.e., \((\alpha_{1,2},\alpha_{3,4})\neq 0\).
Since the two excited atoms reside in the same resonator, they couple strongly to the doublon continuum, in stark contrast to the type-I case.

In type-I pairs, the excitations reside almost entirely on the atoms, and the photonic population in the waveguide is negligible.
However, for type-II pairs we find a small but discernible photonic component at eigenstate \(q=78\) [Fig.~\ref{energy}(c)]:
photons are localized near the atoms yet display an irregular spatial profile (blue histogram).
To quantify the photon distribution, we evaluate $\langle a_j^\dagger a_j\rangle=P_{\rm{one}}(j)+P_{\rm{two}}(j)$, where
\begin{align}
P_{\rm{one}}(j)&=\langle G,{\rm vac}|\sum_{n,i}\sum_{n_1,i_1}d_{n,i}^*d_{n_1,i_1}\sigma_n^-a_ia_j^\dagger a_j\sigma_{n_1}^+a_i^\dagger|G,{\rm vac}\rangle\nonumber\\
&=\sum_{n}|d_{n,j}|^2,\\
P_{\rm{two}}(j)&=\langle {\rm vac}|\sum_{m\leq p}\sum_{m_1\leq p_1}c_{m,p}^*c_{m_1,p_1}a_ma_pa_j^\dagger a_ja_{m_1}^\dagger a_{p_1}^\dagger|{\rm vac}\rangle\nonumber\\
&=\sum_{m\neq j}|c_{j,m}|^2+4|c_{j,j}|^2.
\end{align}
which shows that \(\langle a_j^\dagger a_j\rangle\) comprises a one-photon contribution $P_{\rm{one}}$ and a two-photon contribution $P_{\rm{two}}$.
Accordingly, the blue histogram in Fig.~\ref{energy}(c) can be interpreted as the superposition of an exponentially decaying $P_{\rm{one}}$ localized near the coupling sites (orange histogram) and a standing-wave $P_{\rm{two}}$ between the coupling sites (green histogram), as illustrated in the inset panels.
The two-photon probability distribution \(\langle a_i^\dagger a_j^\dagger a_i a_j\rangle\) for this state is plotted in Fig.~\ref{energy}(d), where photons are clearly confined between the coupling sites in the form of doublons.
Since the corresponding eigenenergy lies within the doublon continuum while the photonic component remains spatially localized, we identify this state as a BIDC.
For comparison, in our previous single-excitation BIC, only the standing-wave component appears.
In the present BIDC, the two atomic pairs carry opposite phases, \(\alpha_{1,2}=-\alpha_{3,4}\) [Fig.~\ref{energy}(\(b_6\))].
By contrast, for a doublon state that is {not} a BIDC, the two pairs are in phase, \(\alpha_{1,2}=\alpha_{3,4}\) [Fig.~\ref{energy}(\(b_5\))], and the two-photon probability in Fig.~\ref{energy}(e) extends across the entire waveguide, indicating the absence of spatial binding between the atoms.

\subsection{Effective model}
Having established the key features of the model through numerical analysis, we now proceed to an analytical treatment to develop a more intuitive understanding.
We expand the wave function of the whole structure as
\begin{align}
|\psi(t)\rangle=&e^{-2i\Omega t}\bigg[\sum_{i<j}Ce_{i,j}(t)\sigma_i^+\sigma_j^++\sum_KC_K(t)D_K^\dagger\nonumber\\
&+\sum_{i,k}c_{i,k}(t)\sigma_i^+a_k^\dagger\bigg]|G,{\rm vac}\rangle+|\psi_s(t)\rangle.
\end{align}
where $a_k=\sum_ne^{ikn}a_n^\dagger/\sqrt{N_c}$ and $D_K^\dagger$ is the creation operator for the $K$th doublon mode, i.e., $|\psi_K^D\rangle=D_K^\dagger|\rm vac\rangle$.
Since we consider the weak coupling condition $g_i\ll-2J-\Omega$, the two-photon scattering state $|\psi_s(t)\rangle$ makes a negligible contribution to the atomic dynamics and will be omitted in what follows~\cite{wx,wangprl}. It is known that there is a wide gap between the single photon scattering continuum and the single atom transition frequency, i.e., $\delta_{k}=-2J\cos(k)-\Omega\gg g_i/\sqrt{N_c}$, the single photon states can be eliminated using perturbation theory. According to the Sch\"odinger equation $H|\psi(t)\rangle=i\partial_t|\psi(t)\rangle$, we obtain the effective coupling between two type-II pairs and the doublon continuum~(see the Appendix~\ref{A} for details),
\begin{align}
i\dot{C}e_{1,2}=&-\frac{g_1^2}{J\sqrt{N_c}}\sum_Kf_K^*(0)e^{iKN_1}C_K,\nonumber\\
i\dot{C}e_{3,4}=&-\frac{g_3^2}{J\sqrt{N_c}}\sum_Kf_K^*(0)e^{iKN_2}C_K,\nonumber\\
i\dot{C_K}=&\Delta_KC_K-\frac{g_1^2}{J\sqrt{N_c}}f_K(0)e^{-iKN_1}Ce_{1,2}\nonumber\\
&-\frac{g_3^2}{J\sqrt{N_c}}f_K(0)e^{-iKN_2}Ce_{3,4}.
\label{DY}
\end{align}
where $\Delta_K=\mathcal{E}_K-2\Omega$. The function $f_K(0)$ contributes to the coupling strength between type-II pairs and the $K$th doublon mode,
its expression is
\begin{align}
f_K(0)=\frac{2\sqrt{2}J}{N_c}\sum_k\frac{\sinh(\lambda_K^{-1})\sqrt{\tanh(\lambda_K^{-1})}}{(\omega_k-\Omega)[\cosh(\lambda_K^{-1})-\cos(k-\frac{K}{2})]}.
\end{align}
According to the Eq.~(\ref{DY}), the effective Hamiltonian of the whole system is
\begin{eqnarray}
H_{\rm {eff}}&=&\Omega\sum_{i=1}^{4}\sigma_{i}^{+}\sigma_{i}^{-}+\sum_{K}\mathcal{E}_KD_K^\dagger D_K\nonumber\\
&&-\frac{g_1^2}{J\sqrt{N_c}}\sum_{K} [f_K(0)e^{-iKN_1}\mathcal{A}_1D_K^{\dagger}+{\rm H.c.}]\nonumber\\
&&-\frac{g_3^2}{J\sqrt{N_c}}\sum_{K} [f_K(0)e^{-iKN_2}\mathcal{A}_2D_K^{\dagger}+{\rm H.c.}].\label{eff}
\end{eqnarray}
where $\mathcal{A}_1=\sigma_{1}^{-}\sigma_{2}^{-}$ and $\mathcal{A}_2=\sigma_{3}^{-}\sigma_{4}^{-}$.
Based on the effective Hamiltonian, we proceed to analyze the eigenstates in the doublon continuum. The corresponding normalized eigenstates are set as
\begin{align}
|\phi_1\rangle=\left(\alpha_1\mathcal{A}_1^\dagger+\alpha_2\mathcal{A}_2^\dagger+\sum_KC_KD_K^\dagger\right)|G,\rm{vac}\rangle.
\end{align}
By applying $H_{\rm{eff}}|\phi_1\rangle=E|\phi_1\rangle$, we obtain the coefficient relationship of the eigenfunction as
\begin{align}
2\Omega-E&=\sum_K\frac{g_1^4f_K^2(0)}{J^2{N_c}}\frac{e^{iK_0(N_2-N_1)}e^{iK(N_1-N_2)}-1}{E-\mathcal{E}_K},\label{eig1}\\
\frac{C_K}{\alpha_1}&=\frac{g_1^2f_K(0)}{J\sqrt{N_c}}\frac{e^{iK_0(N_2-N_1)}e^{-iKN_2} - e^{-iKN_1}}{E-\mathcal{E}_K},\label{eig2}
\end{align}
where wave vector $K_0$ satisfy $2\Omega=\mathcal{E}_{K_0}$. Numerical analysis of Eq.~(\ref{eig1}) reveals the presence of a solution at $E = 2\Omega$ with $\Omega\simeq \Omega_1=\mathcal{E}_{\pi/2}/2$ and $\Delta N\equiv N_2-N_1=8$.
According to the eigenstate in the real space $|\phi_1^r\rangle=\left(\alpha_1\mathcal{A}_1^\dagger+\alpha_2\mathcal{A}_2^\dagger+\sum_{m,n}C_{m,n}a_m^\dagger a_n^\dagger\right)|G,\rm{vac}\rangle$, we derive the photon distribution for the eigenstate of eigenvalue $2\Omega$, that is
\begin{align}
\frac{C_{m,n}}{\alpha_1}=&\frac{1}{\sqrt{2N_c}}\sum_K \frac{C_K}{\alpha_1}e^{iK\frac{m+n}{2}}\psi_{K}(m-n)\nonumber\\
=&\sum_K\frac{g_1^2f_K(0)e^{iK(m+n)/2}\psi_K(m-n)}{J\sqrt{2}N_c}\nonumber\\
&\frac{e^{iK_0(N_2-N_1)}e^{-iKN_2} - e^{-iKN_1}}{2\Omega-\mathcal{E}_K}.\label{eig3}
\end{align}
In our analytical treatment, retaining only the two-photon processes, and the photon distribution is evaluated as
\begin{align}
\langle a_j^\dagger a_j\rangle&=P_{\rm{two}}(j)\nonumber\\
&=\sum_{n',m'}\sum_{n,m}C_{n',m'}^*C_{n,m}\langle {\rm{vac}}|a_{n'}a_{m'}a_j^\dagger a_ja_n^\dagger a_m^\dagger|{\rm{vac}}\rangle\nonumber\\
&=4\sum_n|C_{j,n}|^2.\label{cmn}
\end{align}
The photon distribution, obtained by numerically solving Eq.~(\ref{eig3}) under the eigenstate normalization condition, is shown as purple circles in Fig.~\ref{energy}(c). The excellent agreement with the full numerical results validates the analytical treatment.

\section{entangled state preparation}
The type-II pairs within the above BIDC are entangled and exhibit large excitation probabilities.
In this section, we harness the BIDC to prepare four-atom entangled states when {they are} spatially separated.
Under the Markovian approximation that the CRW is treated as the environment, the dynamics of atoms is governed by the master equation~(details in Appendix~\ref{B})
\begin{align}
\frac{d\rho}{dt}=&-i[H_1,\rho]+\gamma_1L_{[\mathcal{A}_1^\dagger,\mathcal{A}_1]}\rho+\gamma_2L_{[\mathcal{A}_2^\dagger,\mathcal{A}_2]}\rho\nonumber\\
&+\gamma_cL_{[\mathcal{A}_1^\dagger,\mathcal{A}_2]}\rho+\gamma_cL_{[\mathcal{A}_2^\dagger,\mathcal{A}_1]}\rho.\label{master}
\end{align}
where $L_{[O_1,O_2]}\rho=2O_2\rho O_1-\rho O_1O_2-O_1O_2\rho$.
Here we apply a resonant driving field, described by the Hamiltonian $H_1=\eta\Theta(t_0-t)(\mathcal{A}_1+\mathcal{A}_1^\dagger)$ with $\eta$ and $t_0$ being the driving strength and duration respectively, $\Theta(\cdot)$ is the Heaviside step function. This two qubit driving is feasible in superconducting circuits~\cite{wx,pump1,pump2}. The parameters are obtained as
$\gamma_1=g_1^4f_{K_0}^2(0)/v_g(K_0)$ and $\gamma_2=g_3^4f_{K_0}^2(0)/v_g(K_0)$ being the individual decay rates for two type-II pairs, where $v_g(K)$ is the group velocity of the doublon.  $\gamma_c=g_1^2g_3^2f_{K_0}^2(0)\cos[K_0(N_1-N_2)]/v_g(K_0)$ is their collective decay rate.
When $\cos[K_0(N_1-N_2)]=\pm1$, Eq.~(\ref{master}) can be simplified as
\begin{align}
\frac{d\rho}{dt}=&-i[H_1,\rho]+\gamma'L_{[\mathcal{K}^\dagger,\mathcal{K}]}\rho.\label{master1}
\end{align}
where $\gamma'=\gamma_1+\gamma_2$ and $\mathcal{K}^\dagger=(g_1^2\mathcal A_1^\dagger\pm g_3^2\mathcal A_2^\dagger)/\sqrt{g_1^4+g_3^4}$. Subsequently, we find a dark state
\begin{align}
|D\rangle=\frac{1}{{\sqrt{g_1^4+g_3^4}}}(g_3^2\mathcal A_1^\dagger\mp g_1^2\mathcal A_2^\dagger)|G\rangle.\label{dark}
\end{align}
which is orthogonal to the bright state $|B\rangle=\mathcal K^\dagger|G\rangle$.
For \(\Delta N=8\) and \(g_1=g_3\), the dark state reduces to
$|D\rangle=(\mathcal A_1^\dagger -\mathcal A_2^\dagger)|G\rangle/{\sqrt{2}}$
in one-to-one correspondence with the eigenstate highlighted in Fig.~\ref{energy}(\(b_6\)).
There is thus a precise mapping between the BIDC of the {closed} system and the dark state of the {open} system, a relation also noted in other platforms~\cite{sp2,an16}.
In what follows, we exploit this BIDC-dark state correspondence to prepare high-fidelity atomic entanglement.
Specifically, we fix \(\Delta N=8\) and choose suitable atom-waveguide couplings so that, in the two-excitation manifold, the BIDC coincides with the dark state,
targeting the state
$|\phi\rangle=(|eegg\rangle-|ggee\rangle)/\sqrt{2}$.

\begin{figure}[t]
\centering
\includegraphics[width=4.2cm]{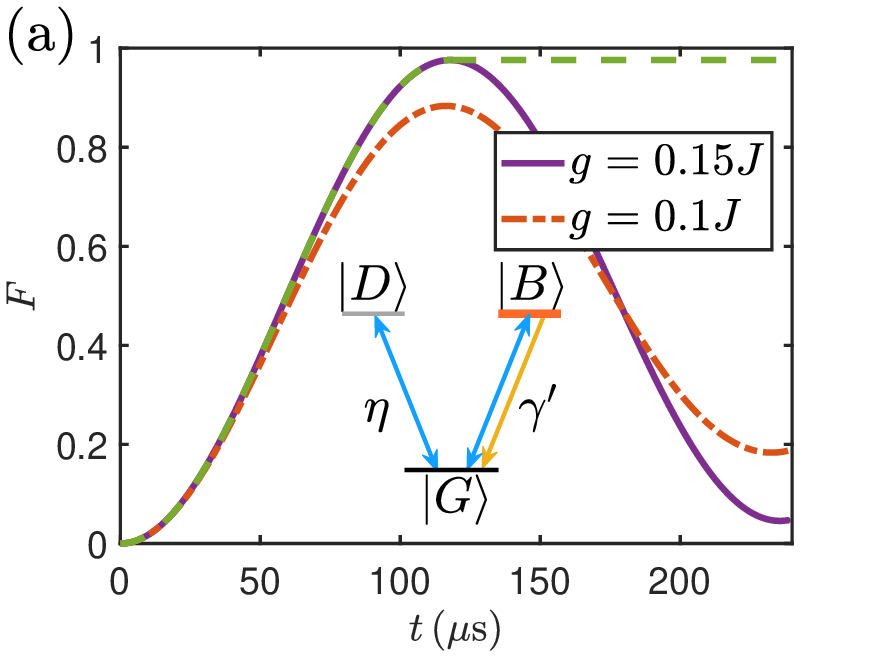}
\includegraphics[width=4.2cm]{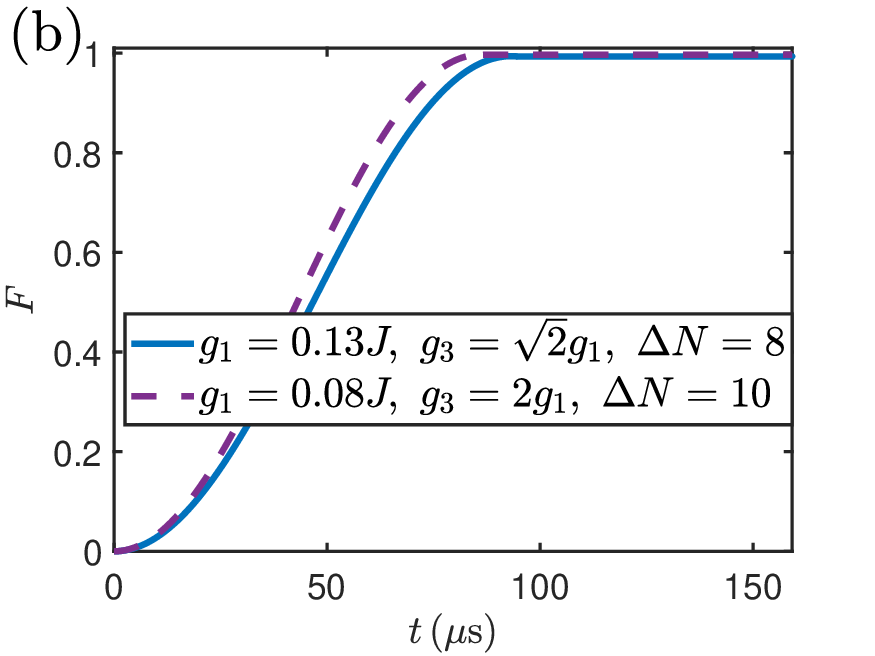}
\caption{(a)~The fidelity of $|D\rangle=(\mathcal A_1^\dagger -\mathcal A_2^\dagger)|G\rangle/{\sqrt{2}}$, with $g_1=g_3=g$ and $\Delta N=8$. The purple solid and red dot-dashed curves correspond to $t_0\rightarrow\infty$. The green dashed curve, with a driving time \(t_0\simeq 118\,{\rm \mu s}\), is otherwise identical to the purple solid curve. (b)~The fidelity of $|D\rangle=(2\mathcal A_1^\dagger -\mathcal A_2^\dagger)|G\rangle/{\sqrt{5}}$~(the blue solid curve) and $|D\rangle=(4\mathcal A_1^\dagger +\mathcal A_2^\dagger)|G\rangle/{\sqrt{17}}$~(the purple dashed curve). The other parameters are set as $2\Omega=\mathcal E_{\pi/2}$ and $\eta=3\times10^{-5}J$, $J/(2\pi)=100\,{\rm MHz}$.}\label{preparation}
\end{figure}

By numerically solving the master equation~(\ref{master1}) from the initial ground state of all atoms, we evaluate the state fidelity with respect to the target entangled state $F=\langle \phi|\rho|\phi\rangle$.
For long driving durations \(t_0\), Fig.~\ref{preparation}(a) displays \(F(t)\) for several atom-waveguide coupling strengths (purple solid and red dot-dashed).
The fidelity first climbs to a maximum and subsequently decreases with time.
As discussed above, the peak coincides with the transient population of the BIDC; switching off the driving at this moment freezes the system into a steady entangled state.
Choosing \(t_0\simeq 118\,{\rm \mu s}\), the green dashed curve in Fig.~\ref{preparation}(a) shows a stabilized fidelity \(F\simeq 0.97\) over long times.
We also find that the peak fidelity increases with stronger atom-waveguide coupling and the underlying physics can be explained as what follows.

Physically, the doublon continuum in the waveguide acts as an effective environment that mediates {only} two-photon processes for the atoms.
Under weak driving, states with more than two excitations can be neglected, and the four-atom manifold reduces to an effective \(V\)-type three-level structure, as sketched in the inset of Fig.~\ref{preparation}(a):
the driving (blue arrows) induces \(\ket{G}\!\leftrightarrow\!\ket{D}\) and \(\ket{G}\!\leftrightarrow\!\ket{B}\) transitions,
whereas the system-environment coupling produces dissipation {only} from the bright state \(\ket{B}\) (yellow arrow) but not from the dark state \(\ket{D}\).
The corresponding decay rate scales as \(g^4\) and enters the master equation as \(\gamma'\).
For stronger atom-waveguide coupling, e.g., \(g=0.15J\), the bright state \(\ket{B}\) decays more rapidly than for \(g=0.1J\),
so the dynamics is effectively a Rabi oscillation predominantly between \(\ket{G}\) and \(\ket{D}\).
At the optimal time \(t_0\), the population is concentrated in \(\ket{D}\) (which coincides with the target state),
and turning off the driving for \(t>t_0\) yields high fidelity.
In contrast, for weaker \(g\), residual population of \(\ket{B}\) is unavoidable at any time,
leading to a lower peak fidelity compared with the strong-coupling case.

Beyond the specific example in Fig.~\ref{preparation}(a), the scheme can be generalized to prepare an arbitrary superposition
\(\ket{\phi}=\alpha\ket{e e g g}+\beta\ket{g g e e}\).
To this end, we choose the couplings such that
$g_3^2/{\sqrt{g_1^4+g_3^4}}=|\alpha|, g_1^2/{\sqrt{g_1^4+g_3^4}}=|\beta|$
and tune \(\Delta N\) to set the relative phase (0 or \(\pi\)),
so that the dark state (which coincides with the BIDC) matches the target \(\ket{\phi}\).
Figure~\ref{preparation}(b) shows the resulting fidelities for
\(\alpha=2/\sqrt{5},\,\beta=-1/\sqrt{5}\) and
\(\alpha=4/\sqrt{17},\,\beta=1/\sqrt{17}\),
obtained by turning off the driving at \(t_0\simeq 94\,{\rm \mu s}\) and
\(t_0\simeq 87\,{\rm \mu s}\), respectively.
In both cases, the fidelity reaches \(F\simeq 0.99\).

\section{entangled state transfer}
\begin{figure}[t]
  \centering
  \includegraphics[width=4.2cm]{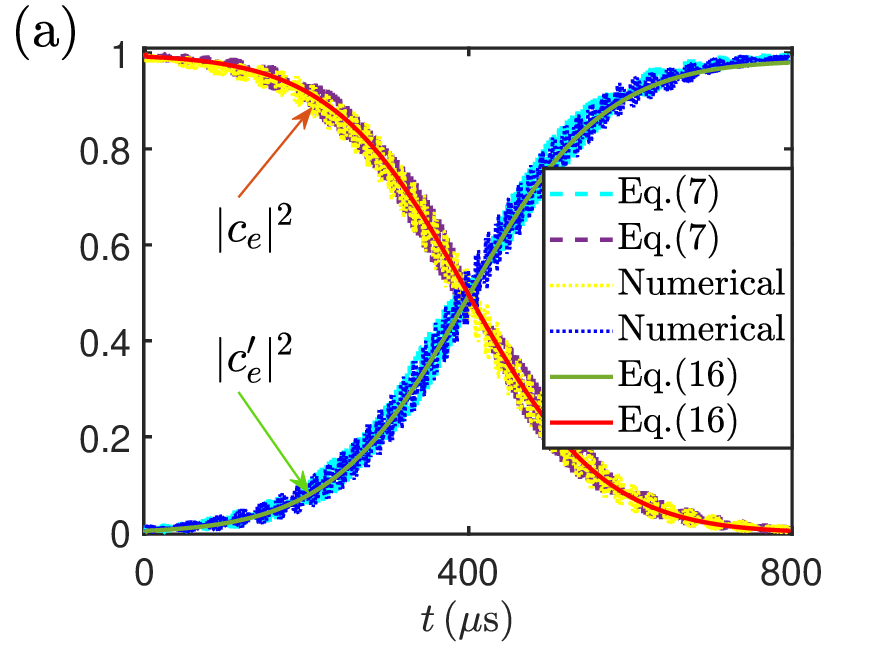}\includegraphics[width=4.2cm]{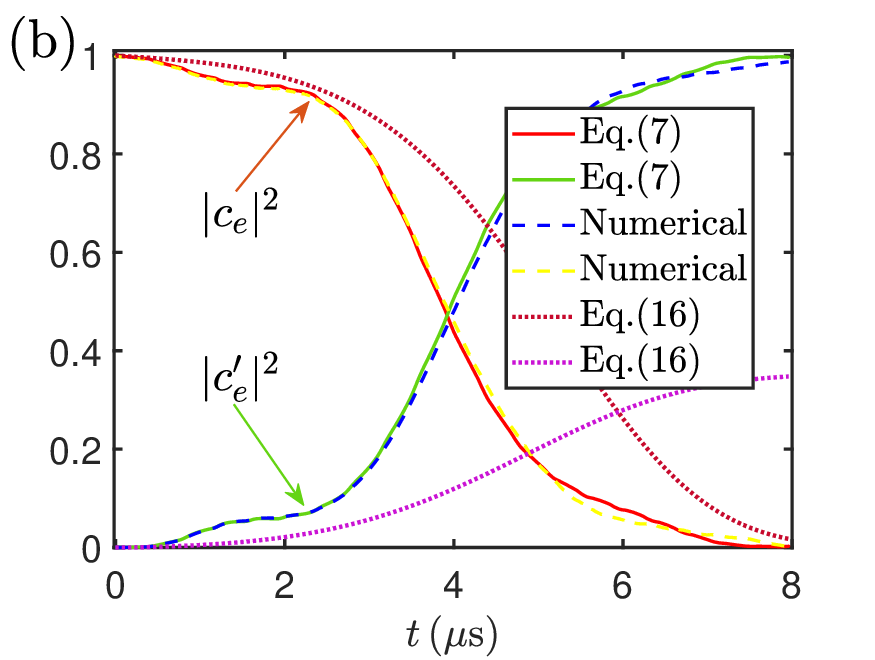}
  \includegraphics[width=4.2cm]{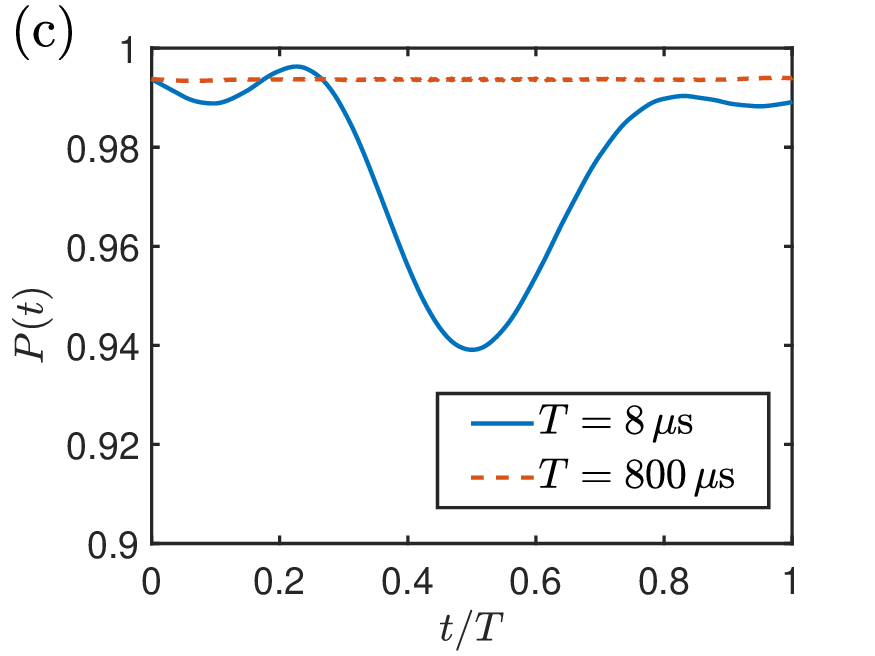}\includegraphics[width=4.2cm]{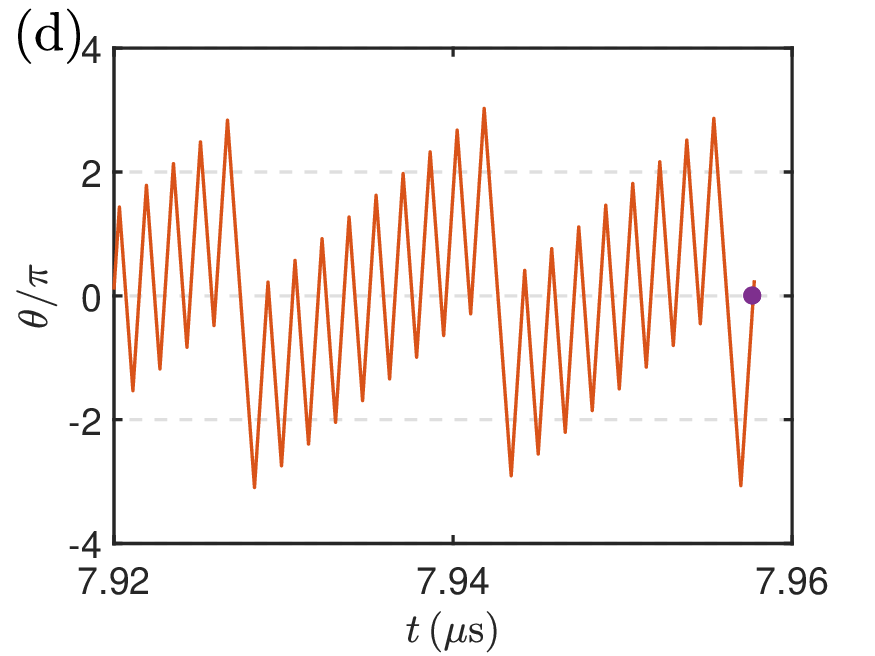}
  \caption{~The excitation probability of atomic pairs in the long time (a) and short time (b)
  state transfer protocol.  (c)~The projection $P(t)$. (d)~The angle of amplitude $\theta$. The parameters are given in the main text. }\label{transition}
\end{figure}
In this section, we will utilize BIDC to transfer a quantum state $c_e|ee\rangle+c_g|gg\rangle$ from one type-II pair to another.
To this end, we prepare the initial state as $|\psi(0)\rangle=\sigma_1^{+}\sigma_2^{+}|G,{\rm vac}\rangle$, at arbitrary moment, the wave function can be expressed as
\begin{align}
|\psi(t)\rangle=&e^{-iHt}|\psi(0)\rangle\nonumber\\=
&\bigg[c_e\sigma_1^{+}\sigma_2^{+}+c_e'e^{i\theta(t)}\sigma_3^{+}\sigma_4^{+}
+\sum_{n,i}d_{n,i}(t)\sigma_n^+a_i^\dagger\nonumber\\
&+\sum_{m,n}c_{i,j}(t)a_i^{\dagger}a_j^\dagger
+\alpha_{1,3}(t)\sigma_1^{+}\sigma_3^{+}+\alpha_{1,4}(t)\sigma_1^{+}\sigma_4^{+}\nonumber\\
&+\alpha_{2,3}(t)\sigma_2^{+}\sigma_3^{+}+\alpha_{2,4}(t)\sigma_2^{+}\sigma_4^{+}\bigg]|G,{\rm vac}\rangle,\label{dou}
\end{align}
with initial condition $c_e(0)=1$. Choosing appropriate time $T$, which satisfies $|c_e'(T)|=|c_e(0)|$ and $\theta(T)=0$, we accomplish an entangled state transfer process, i.e.,  $\mathcal{N}(c_e|ee\rangle_{1,2}+c_g|gg\rangle_{1,2})
\otimes|gg\rangle_{3,4}\rightarrow|gg\rangle_{1,2}\otimes
\mathcal{N}(c_e|ee\rangle_{3,4}+c_g|gg\rangle_{3,4})$.
Here, we note that, the ground state $|gggg\rangle$ will not evolve governed by the Hamiltonian $H$. Therefore, $c_e$ and $c_g$ can be chosen freely, allowed by the normalized constance $\mathcal{N}$. In other words, we here
achieve an arbitrary state transfer.

As the first step, we focus on the transfer of the probability amplitudes of the two type-II pairs  via the BIDC of the atom-waveguide system. Actually, the atomic counterpart of the BIDC coincides  the dark state \(\ket{D}\) (similar mechanism can be also found in Refs.~\cite{sp2,an16}) in Eq.~(\ref{dark}), which depends on the coupling strengths. Therefore, by {adiabatically} varying the couplings \(g_1\) and \(g_3\) so that the system remains on the BIDC throughout the evolution, one can coherently shuttle the excitation probability from one type-II pair to the other.
To validate this mechanism, we simulate the  dynamics of $|c'_e(t)|$ in Eq.~(\ref{dou}) under the ramp
$g_1=g_2=(0.15t/T+0.05)J$ and $g_3=g_4=(-0.15t/T+0.2)J$,
and a total QST duration \(T=800\,{\rm\mu s}\); the resulting population transfer is shown in Fig.~\ref{transition}(a). The excellent agreement among (i) direct exact numerics, (ii) the single-photon eliminated dynamics of Eq.~(\ref{DY}), and (iii) the master equation treatment of Eq.~(\ref{master1}) [where $|c_e|^2={\rm Tr}(\mathcal{A}_1^\dagger\mathcal{A}_1\rho)$] confirms the reliability of the approach.
While highly faithful, the required time is long.
Reducing the duration by two orders of magnitude to \(T=8\,{\rm\mu s}\) still yields comparable amplitude transfer, as shown in Fig.~\ref{transition}(b).
However, in this faster protocol, the Markovian approximation deviates noticeably from exact numerics and the single-photon elimination result, while the latter two approaches show a good agreement.

To elucidate the breakdown of the master equation description, we compare the dynamics for the slow
(\(T=800\,{\rm\mu s}\))and fast ( (\(T=8\,{\rm\mu s}\)) QST protocols in Fig.~\ref{transition}(c).
Specifically, we monitor the instantaneous overlap with the BIDC,
$P(t)=|\langle\psi(t)|\rm{BIDC}\rangle|^2$,
throughout the transfer process.
For the slow (adiabatic) protocol, the dashed orange curve shows \(P(t)\gtrsim 0.99\) at all times, indicating that the evolution faithfully follows the BIDC with negligible population of other states (i.e., photons are scarcely excited).
In other words, the waveguide which acts as the environment remains essentially unchanged, validating the Markovian approximation, consistent with Fig.~\ref{transition}(a).
In contrast, for the fast protocol, the system temporarily departs from the BIDC and tunnels into scattering states
(solid curve, with \(P(t)\approx 0.94\) at the nadir), before relaxing back toward the BIDC as the evolution proceeds.
In this nonadiabatic regime, the waveguide cannot be treated as a memoryless reservoir, and the Markovian approximation fails, in agreement with Fig.~\ref{transition}(b).

Except for the amplitude dynamics, verifying the relative phase is essential for a faithful QST.
To this end, Fig.~\ref{transition}(d) shows the temporal evolution of \(\theta(t)\) in Eq.~(\ref{dou}).
The phase \(\theta(t)\) exhibits oscillations and crosses zero (purple filled circle), at which point the transferred amplitude also reaches the expected value, \(|c'_e|^2\simeq 1\) [cf. Fig.~\ref{transition}(b)].
At this moment, the state has been successfully transferred from the type-II pair in resonator \(N_1\) to the pair in resonator \(N_2\), completing the QST with both amplitude and phase preserved. Since $c_g$ can be arbitrary, we now realize the transfer of any entangled state $c_e|ee\rangle+c_g|gg\rangle$.

\section{Conclusion}
We have introduced a waveguide-QED platform where four two-level atoms couple to the CRW with strong on-site interactions, realizing a BIDC.
In the two-excitation sector, the CRW hosts both a two-photon scattering continuum and an interaction-induced doublon continuum.
We identified a family of quasi-degenerate eigenstates with nearly two atomic excitations; among them, a genuine BIDC emerges.
Crucially, we established a one-to-one correspondence between this closed-system BIDC and the dark state of the corresponding open system, enabling dissipative preparation of remote multi-atom entanglement with high fidelity.
Furthermore, we proposed and validated a BIDC-enabled QST protocol that shuttles excitations between remote type-II pairs by ramping the atom-waveguide couplings.
We verified that faithful QST preserves both the population modulus and phase of the logical state, with the relative phase \(\theta\) synchronized to the amplitude peak.

Superconducting qubits can be coupled to microwave resonator arrays, with embedded Josephson junctions yielding strong intrinsic nonlinearity. In such a system where $g\ll J$, the on-site interaction $U/(2\pi)\simeq100-800\,{\rm MHz}$, all of which are achievable with state-of-the-art experimental techniques~\cite{U1,U2,J1,U3}.
When $J/(2\pi)=100\,{\rm MHz}$, the time required for entangled state preparation is about $100\,{\rm \mu s}$, and the duration for the fast QST protocols is $8\,{\rm \mu s}$; both are faster than the decay time $T_1=500\,{\rm \mu s}$ of the bare qubits~\cite{LS1}. In our proposal, we assume a linear time dependence of the atom--waveguide coupling strength during the QST process. Physically, it is the time-reversal relation between $g_1(t)$ and $g_3(t)$, rather than their specific functional forms, that plays the essential role, which further relaxes the experimental requirements and enhances the feasibility of the scheme.

Our results establish BIDC as a versatile resource for scalable remote entanglement generation and coherent state transfer in strongly correlated photonic media.

{\bf Note:} Just in the submission of this manuscript, we note that the bound states in the doublon continuum was recently investigated~\cite{BIDC1}, but realized by the giant atoms, which interact with coupled resonator waveguide.

\section*{Acknowledgments}

This work is supported by the Natural Science Foundation of China (Grants Nos.~12375010 and 12174303), the Science Foundation of the Education Department of Jilin Province (No.~JJKH20250301KJ) and Science and Technology Development
Project of Jilin Province (Grant No. 20230101357JC).

\section*{DATA AVAILABILITY}

The data that support the findings of this article are not
publicly available. The data are available from the authors
upon reasonable request.

\appendix
\addcontentsline{toc}{section}{Appendices}\markboth{APPENDICES}{}
\begin{subappendices}
\begin{widetext}
\section{The effective Hamiltonian}\label{A}
In our model, four atoms are coupled to the CRW, with the first and second atoms located at the $N_1$th site and the third and fourth ones at the $N_2$ site.
The Hamiltonian of the whole structure is $H=H_A+H_c+H_I$,
\begin{eqnarray}
H_A&=&\sum_i\Omega_i \sigma_i^+\sigma_i^-,\nonumber\\
H_c&=&\sum_n \omega_ca_n^\dagger a_n-\frac{U}{2}\,a_n^\dagger a_n^\dagger a_n a_n
- J\left(a_n^\dagger a_{n+1}+\rm{H.c.}\right),\nonumber\\
H_I&=&\sum_{i=1}^{4} g_i(\sigma_i^+a_{n_i}+{\rm H.c.}).
\end{eqnarray}
Here, the on-site interaction in the CRW supports the doublon energy spectrum, with its eigenenergies and eigenvectors given by~\cite{doublon1,doublon2}
\begin{align}
\mathcal{E}_K=&2\omega_c-\sqrt{U^2+16J^2\cos(K/2)^2},\nonumber\\
|\psi_K^D\rangle=&D_K^\dagger|vac\rangle=\frac{1}{\sqrt{2}}\frac{1}{\sqrt{N_c}}\sum_{m,n}e^{iK\frac{m+n}{2}}\psi_{K}(m-n)a_m^\dagger a_n^\dagger|vac\rangle.
\end{align}
where $\psi_{K}(m-n)=\sqrt{\tanh(\lambda_K^{-1})}e^{-|m-n|/\lambda_K}$ and $\lambda_K^{-1}={\rm asinh}(\frac{U}{4J\cos(K/2)})$.
We assume the four atoms have nearly identical transition frequencies, with $\Omega_i\approx\Omega$.
We further set the atomic transition frequency to be greatly detuned from the single photon scattering continuum~($-2J-\Omega\gg g_i$) but such that twice this frequency is resonant with the doublon continuum~($2\Omega\in\mathcal{E}_K^D$).
In the double excitation subspace, the wave function can be assumed to be
\begin{align}
|\psi(t)\rangle=e^{-2i\Omega t}\left[\sum_{i<j}Ce_{i,j}(t)\sigma_i^+\sigma_j^++\sum_KC_K(t)D_K^\dagger+\sum_{i,k}c_{i,k}(t)\sigma_i^+a_k^\dagger\right]|G,{\rm vac}\rangle+|\psi_s(t)\rangle.
\end{align}
Since we consider the weak coupling condition $g_i\ll-2J-\Omega$, the two-photon scattering state $|\psi_s(t)\rangle$ makes a negligible contribution to the atomic dynamics and will be omitted in what follows. Therefore, the dynamics equations for the wave function amplitude (governed by Sch\"{o}dinger equation) are
{\begin{align}
 &i\dot{c}_{i,k}(t)=(\omega_k+\Omega_i-2\Omega){c}_{i,k}(t)+\sum_{K}g_iC_K(t)M(k,n_i,K)+\frac{1}{\sqrt{N_c}}\sum_{j\neq i}g_{j}Ce_{[i,j]}(t)e^{-ikn_{j}},\label{dy1}\\
  &i\dot{Ce}_{i,j}(t)=(\Omega_i+\Omega_j-2\Omega){Ce}_{i,j}(t)+\frac{1}{\sqrt{N_c}}\sum_k\left(g_j{c}_{i,k}(t)e^{ikn_{j}}+g_i{c}_{j,k}(t)e^{ikn_{i}}\right),\label{dy2}\\
 &i\dot{C}_K(t)=(\mathcal{E}_K-2\Omega){C}_K(t)+\sum_{i,k}g_ic_{i,k}M^*(k,n_i,K).\label{DY1}
 \end{align}}
 where $a_k^\dagger=\sum_je^{ikj}a_j^\dagger/\sqrt{N_c}$ and $M(k,n,K)=\langle {\rm vac}|a_ka_nD_K^\dagger|{\rm vac}\rangle=\sqrt{2}\sum_me^{-ikm}e^{iK(n+m)/2}\psi_K(n-m)/N_c$. Symbol $[i,j]$ permutes $i$ and $j$ such that the smaller value comes first and the larger one comes last, that is, $[i,j]=\min\{i,j\},\max\{i,j\}$.
We set $\delta_{k,i}=\omega_k+\Omega_i-2\Omega$ and $\Delta_K=\mathcal{E}_K-2\Omega$.
Adiabatically eliminating the amplitudes $c_{ik}$ by setting $\dot{c}_{ik}=0$, Eq.~(\ref{dy1}) can be simplified as
{\begin{align}
{c}_{i,k}(t)=\frac{-1}{\delta_{k,i}}\left[\sum_{K}g_iC_K(t)M(k,n_i,K)+\frac{1}{\sqrt{N_c}}\sum_{j\neq i}g_{j}Ce_{[i,j]}(t)e^{-ikn_{j}}\right].
 \end{align}}

Taking above equation into Eqs.~(\ref{dy2},\ref{DY1}), we can get
\begin{align}
i\dot{Ce}_{i,j}=&(\Omega_i+\Omega_j-2\Omega)Ce_{i,j}(t)\nonumber\\
&-\frac{g_ig_j}{J\sqrt{N_c}}\sum_{K}f_K^*(n_i,n_j)e^{iK(n_i+n_j)/2}C_K(t)\nonumber\\
&-\sum_{j_1\neq i}\frac{g_jg_{j_1}}{J}G(n_j-n_{j_1})Ce_{[i,j_1]}
-\sum_{j_1\neq j}\frac{g_ig_{j_1}}{J}G(n_i-n_{j_1})Ce_{[j,j_1]},\nonumber\\
i\dot{C_K}(t)=&\Delta_KC_K(t)-\sum_{i<j}\frac{g_ig_j}{J\sqrt{N_c}}f_K(n_i,n_j)e^{-iK(n_i+n_j)/2}Ce_{i,j}(t).\label{DY2}
\end{align}
Here, we ignore the interaction between the different doublon mode and take $\delta_{k,i}=\delta_k=\omega_k-\Omega$.
\begin{figure}
  \centering
  \includegraphics[width=7cm]{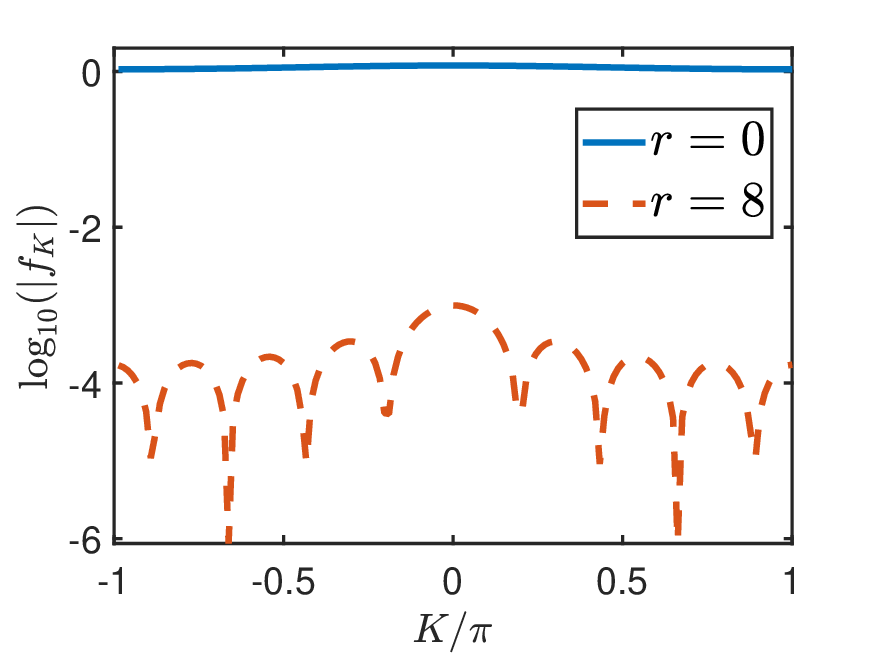}\\
  \caption{~The interaction strength $|f_K(r)|$.}\label{fk}
\end{figure}
The interaction strength between atom pair and doublon is
{\begin{align}
f_K(r_{i,j})=&\sum_kJe^{iK(n_i+n_j)/2}\left[\frac{e^{-ikn_j}}{\delta_{k}}M^*(k,n_i,K)+\frac{e^{-ikn_i}}{\delta_{k}}M^*(k,n_j,K)\right]\nonumber\\
=&\frac{2\sqrt{2}J}{N_c}\sum_k\frac{L_{k,K}\cos \left[(k-K/2)r_{i,j}\right] }{\delta_k},\\
L_{k,K}=&\frac{\sinh(\lambda_K^{-1})\sqrt{\tanh(\lambda_K^{-1})}}{(\cosh(\lambda_K^{-1})-\cos(k-\frac{K}{2}))}.
\end{align}}
where $r_{i,j}=n_i-n_j$. As shown in Fig.~\ref{fk}, the interaction between type-I pairs and the doublon~($f_K(8)$ ) is much weaker than that of type-II~($f_K(0)$).
Because the effective coupling strength between type-I pairs and the doublon continuum is sufficiently small, the atoms in the eigenstates represented by the blue histograms in Fig.~\ref{energy}(b) do not decay.
Next, we neglect the interaction between type-I pairs and the doublon in the case of $\Delta N=8$, which is discussed in the main text.

The single photon process can mediate the interaction between different atomic pairs, and the interaction strength is
{\begin{align}
G(n)=&\frac{J}{N_c}\sum_k\frac{e^{ikn}}{\omega_k-\Omega}=\frac{J}{2\pi}\int_{-\pi}^{\pi}dk \frac{e^{ikn}}{\omega_k-\Omega}=\frac{J(-1)^{|n|}}{\sqrt{\Omega^2-4J^2}}\left[\frac{\Omega}{2J}+\sqrt{\frac{\Omega^2}{4J^2}-1}\right]^{|n|}
=\frac{Je^{-a|n|}}{\sqrt{\Omega^2-4J^2}},\nonumber\\
a=&\ln\left[\frac{-\Omega}{2J}+\sqrt{\frac{\Omega^2}{4J^2}-1}\right].\label{G}
\end{align}}
Remarkably, $\Omega<-2J$, so $a>0$ and $a$ increases as $\Omega$ moves away from the single photon scattering continuum. For $U=6J$ and $2\Omega\simeq\mathcal{E}_{\pi/2}$, $G(0)\gg G(8)$, we ignore the interaction between the different atom pairs but keep the Stark shift. Eq.~(\ref{DY2}) can be simplified as
{\begin{align}
i\dot{Ce}_{1,2}=&\left(\Omega_1+\Omega_2-2\Omega-\frac{g_2^2}{J}G^{(1)}(0)-\frac{g_1^2}{J}G^{(2)}(0)\right)Ce_{1,2}(t)-\frac{g_1g_2}{J\sqrt{N_c}}\sum_{K}f_K^*(0)e^{iKN_1}C_K(t),\nonumber\\
i\dot{Ce}_{3,4}=&\left(\Omega_3+\Omega_4-2\Omega-\frac{g_4^2}{J}G^{(3)}(0)-\frac{g_3^2}{J}G^{(4)}(0)\right)Ce_{3,4}(t)-\frac{g_3g_4}{J\sqrt{N_c}}\sum_{K}f_K^*(0)e^{iKN_2}C_K(t),\nonumber\\
i\dot{C_K}(t)=&\Delta_KC_K(t)-\frac{g_1g_2}{J\sqrt{N_c}}f_K(0)e^{-iKN_1}Ce_{1,2}(t)-\frac{g_3g_4}{J\sqrt{N_c}}f_K(0)e^{-iKN_2}Ce_{3,4}(t).\label{DY3}
\end{align}}
 We set $g_1=g_2$, $g_3=g_4$, $\Omega_1=\Omega_2$ and $\Omega_3=\Omega_4$ and take
\begin{align}
\Omega_1=\Omega+\frac{g_1^2}{\sqrt{\Omega^2-4J^2}},\nonumber\\
\Omega_3=\Omega+\frac{g_3^2}{\sqrt{\Omega^2-4J^2}}.
\end{align}
Eqs.~(\ref{DY3}) can be simplified as

{\begin{align}
i\dot{Ce}_{1,2}=&-\frac{g_1^2}{J\sqrt{N_c}}\sum_{K}f_K^*(0)e^{iKN_1}C_K(t),\nonumber\\
i\dot{Ce}_{3,4}=&-\frac{g_3^2}{J\sqrt{N_c}}\sum_{K}f_K^*(0)e^{iKN_2}C_K(t),\nonumber\\
i\dot{C_K}(t)=&\Delta_KC_K(t)-\frac{g_1^2}{J\sqrt{N_c}}f_K(0)e^{-iKN_1}Ce_{1,2}(t)-\frac{g_3^2}{J\sqrt{N_c}}f_K(0)e^{-iKN_2}Ce_{3,4}(t).\label{DY4}
\end{align}}
According to Eqs.~(\ref{DY4}), we can get the effective Hamiltonian of the whole structure
{\begin{align}
H_{eff}=&\Omega\sum_{i=1}^{4}\sigma_{i}^{+}\sigma_{i}^{-}+\sum_{K}\mathcal{E}_KD_K^\dagger D_K\nonumber\\
&-\frac{g_1^2}{J\sqrt{N_c}}\sum_{K} [f_K(0)e^{-iKN_1}\sigma_{1}^{-}\sigma_{2}^{-}D_K^{\dagger}+{\rm H.c.}]-\frac{g_3^2}{J\sqrt{N_c}}\sum_{K} [f_K(0)e^{-iKN_2}\sigma_{3}^{-}\sigma_{4}^{-}D_K^{\dagger}+{\rm H.c.}].\label{Heff1}
\end{align}}

The eigenstate in the two excitation subspace can be written as
\begin{align}
|\phi_1\rangle=\left(\alpha_1\mathcal{A}_1^\dagger+\alpha_2\mathcal{A}_2^\dagger+\sum_KC_KD_K^\dagger\right)|G,\rm{vac}\rangle.
\end{align}
Substitution of the above expression into the eigenequation $H_{\rm{eff}}|\phi_1\rangle=E|\phi_1\rangle$ leads to the equations satisfied by the coefficients
\begin{align}
E\alpha_1 &= 2\Omega \alpha_1 - \frac{g_1^2}{J\sqrt{N_c}}\sum_K f_K(0)e^{iKN_1}C_K,\label{a1} \\
E\alpha_2 &= 2\Omega \alpha_2 - \frac{g_3^2}{J\sqrt{N_c}}\sum_K f_K(0)e^{iKN_2}C_K, \label{a2}\\
E C_K &= \mathcal{E}_K C_K - \frac{g_1^2}{J\sqrt{N_c}} f_K(0)e^{-iKN_1}\alpha_1 - \frac{g_3^2}{J\sqrt{N_c}} f_K(0)e^{-iKN_2}\alpha_2.\label{ck}
\end{align}
Considering the special case of $E = \mathcal{E}_{K_0}$ and substituting it into Eq.~(\ref{ck}), we obtain the relation between the excitation probability amplitudes of the two atom pair
\begin{equation}
\alpha_2 = -\frac{g_1^2}{g_3^2}e^{iK_0(N_2-N_1)}\alpha_1.\label{a3}
\end{equation}
Using the Eq.~(\ref{a3}), Eqs.~(\ref{a1}-\ref{ck}) can be simplified to
\begin{align}
E-2\Omega&=-\sum_K\frac{g_1^4f_K^2(0)}{J^2{N_c}}\frac{e^{iK_0(N_2-N_1)}e^{iK(N_1-N_2)}-1}{E-\mathcal{E}_K},\label{b1}\\
\frac{C_K}{\alpha_1}&=\frac{g_1^2f_K(0)}{J\sqrt{N_c}}\frac{e^{iK_0(N_2-N_1)}e^{-iKN_2} - e^{-iKN_1}}{E-\mathcal{E}_K}.\label{b2}
\end{align}
Numerical analysis of Eq.~(\ref{b1}) reveals the presence of a solution at $E = 2\Omega$ with $\Omega\simeq \Omega_1=\mathcal{E}_{\pi/2}/2$ and $\Delta N=8$.
Through a Fourier transformation of Eq.~(\ref{b2}), we derive the photon distribution for the eigenstate of eigenvalue $2\Omega$, that is
\begin{align}
\frac{C_{m,n}}{\alpha_1}=&\frac{1}{\sqrt{2N_c}}\sum_K \frac{C_K}{\alpha_1}e^{iK\frac{m+n}{2}}\psi_{K}(m-n)\nonumber \\
=&\sum_K\frac{g_1^2f_K(0)e^{iK(m+n)/2}\psi_K(m-n)}{J\sqrt{2}N_c}\frac{e^{iK_0(N_2-N_1)}e^{-iKN_2} - e^{-iKN_1}}{2\Omega-\mathcal{E}_K}.
\end{align}
where the eigenstate in the real space $|\phi_1^r\rangle=\left(\alpha_1\mathcal{A}_1^\dagger+\alpha_2\mathcal{A}_2^\dagger+\sum_{m,n}C_{m,n}a_m^\dagger a_n^\dagger\right)|G,\rm{vac}\rangle$.

\section{Master equation}\label{B}
Under the Markovian approximation and working in the interaction picture, the formal master equation for open system is
\begin{align}
\dot\rho=-\int_0^\infty d\tau {\rm Tr}_c\left[H_I(t),[H_I(t-\tau),\rho_c\otimes\rho]\right].
\end{align}
According to Eq.~(\ref{Heff1}), the interaction Hamiltonian is
\begin{align}
H_I=-\frac{g_1^2f_K(0)}{J\sqrt{N_c}}\sum_K\left[e^{-iKN_1}e^{i(\mathcal E_K-2\Omega)t}\mathcal{A}_1D_K^++\rm {H.c.}\right]-\frac{g_3^2f_K(0)}{J\sqrt{N_c}}\sum_K\left[e^{-iKN_2}e^{i(\mathcal E_K-2\Omega)t}\mathcal{A}_2D_K^++\rm {H.c.}\right].
\end{align}
where $\mathcal{A}_1=\sigma_1^-\sigma_2^-$ and $\mathcal{A}_2=\sigma_3^-\sigma_4^-$. The master equation tracing out the degree of freedom of the waveguide is then obtained as
\begin{align}
\frac{d\rho}{dt}=&A\left[2\mathcal{A}_1\rho\mathcal{A}_1^\dagger-\rho\mathcal{A}_1^\dagger\mathcal{A}_1-\mathcal{A}_1^\dagger\mathcal{A}_1\rho\right]
+D\left[2\mathcal{A}_2\rho\mathcal{A}_2^\dagger-\rho\mathcal{A}_2^\dagger\mathcal{A}_2-\mathcal{A}_2^\dagger\mathcal{A}_2\rho\right]\nonumber\\
&+(B+C)\mathcal{A}_1\rho\mathcal{A}_2^\dagger-B\rho\mathcal{A}_2^\dagger\mathcal{A}_1-C\mathcal{A}_2^\dagger\mathcal{A}_1\rho\nonumber\\
&+(C+B)\mathcal{A}_2\rho\mathcal{A}_1^\dagger-C\rho\mathcal{A}_1^\dagger\mathcal{A}_2-B\mathcal{A}_1^\dagger\mathcal{A}_2\rho.
\end{align}
where
\begin{align}
A&=\int_0^\infty d\tau\sum_K\frac{g_1^4f_K^2(0)}{J^2N_c}e^{-i(\mathcal E_K-2\Omega)\tau}\nonumber\\
&=\frac{g_1^4}{2\pi J^2}\int_{-\pi}^{\pi}dKf_K^2(0)\int_0^\infty d\tau e^{-i(\mathcal E_K-2\Omega)\tau}\nonumber\\
&=\frac{g_1^4}{2 J^2}\int_{-\pi}^{\pi}dKf_K^2(0)\delta(\mathcal E_K-2\Omega)\nonumber\\
&=\frac{g_1^4}{2 J^2}\left[\int_{0}^{\pi}dKf_K^2(0)\delta(\mathcal E_K-2\Omega)+\int_{-\pi}^{0}dKf_K^2(0)\delta(\mathcal E_K-2\Omega)\right]\nonumber\\
&=\frac{g_1^4}{2 J^2}\left[\int_{\mathcal{E}_{\rm {min}}}^{\mathcal{E}_{\rm {max}}}\frac{d\mathcal E_K}{v_g(K)}f_K^2(0)\delta(\mathcal E_K-2\Omega)+\int_{\mathcal{E}_{\rm {max}}}^{\mathcal{E}_{\rm {min}}}\frac{d\mathcal E_K}{v_g(K)}f_K^2(0)\delta(\mathcal E_K-2\Omega)\right]\nonumber\\
&=\frac{g_1^4}{2 J^2}\left[\frac{f_{K_0}^2(0)}{v_g(K_0)}-\frac{f_{-K_0}^2(0)}{v_g(-K_0)}\right]\nonumber\\
&=\frac{g_1^4}{J^2}\frac{f_{K_0}^2(0)}{v_g(K_0)}.
\end{align}
where $K_0=\pi/2$. We have taken the fact of that $v_g(K)$ is odd function and $f_K(0)$ is even function. Similarly, we can also get other parameters
\begin{align}
B&=\int_0^\infty d\tau\sum_K\frac{g_1^2g_3^2f_K^2(0)}{J^2N_c}e^{iK(N_1-N_2)}e^{-i(\mathcal E_K-2\Omega)\tau}=\frac{g_1^2g_3^2}{J^2}\frac{f_{K_0}^2(0)}{v_g(K_0)}\cos[K_0(N_1-N_2)],\\
C&=\int_0^\infty d\tau\sum_K\frac{g_1^2g_3^2f_K^2(0)}{J^2N_c}e^{iK(N_2-N_1)}e^{-i(\mathcal E_K-2\Omega)\tau}=\frac{g_1^2g_3^2}{J^2}\frac{f_{K_0}^2(0)}{v_g(K_0)}\cos[K_0(N_2-N_1)],\\
D&=\int_0^\infty d\tau\sum_K\frac{g_3^4f_K^2(0)}{J^2N_c}e^{-i(\mathcal E_K-2\Omega)\tau}=\frac{g_3^4}{J^2}\frac{f_{K_0}^2(0)}{v_g(K_0)}.
\end{align}

\end{widetext}
\end{subappendices}

\end{document}